# Abundance Histories for QSO Absorption Line Systems

F. X. Timmes, J. T. Lauroesch & J. W. Truran

Laboratory for Astrophysics and Space Research
and
Department of Astronomy and Astrophysics
and
Enrico Fermi Institute
University of Chicago
Chicago, IL  60637

ABSTRACT

Abundance histories for QSO absorption line systems as a function of redshift are presented for all metals lighter than gallium. Coupling various conventional cosmological models with a simple, first–order model for the chemical evolution of the QSO absorption line systems allows transformation of the observed abundance trends in the Galaxy to abundance histories for the gas in QSO absorption line systems. Comparison of the transformed abundance patterns with the zinc to hydrogen [Zn/H] and silicon to hydrogen [Si/H] ratios observed in damped Lyman-$\alpha$ systems finds agreement over more than two orders of magnitude in abundance with a $\Lambda$=0, $\Omega$=0.2 – 1.0, and $\tau_{\rm delay}$=3 Gyr cosmology, where $\tau_{\rm delay}$ is the time between the start of the Big Bang and the beginning of galaxy formation. Alternative meanings for the required time delay are explored, along with extensions to the simple first–order model. Predictions are made for the abundance histories of elements not yet observed in QSO absorption line systems.

Subject headings: galaxies: abundances – galaxies: evolution – quasars: absorption line

## 1. INTRODUCTION

Gas enriched by nucleosynthetic processes is detected out to redshifts $z \simeq 4$ by the metal absorption lines they produce in the spectra of the background quasi-stellar objects (QSOs). Analyses of the absorption spectra produced by the gas which is suspected to reside in galaxies can uniquely measure a variety of physical parameters associated with the gas: temperature, column density, chemical abundances, grain content, molecular content, and average supernova rates. Redshifts measured in the absorption systems presumably sample the gas at different points in their roughly common evolutionary paths, and the abundance patterns in these systems offer an understanding of their nucleosynthetic evolution. The heavy element abundances found in the QSO absorbers span the metallicity range $10^{-3}$ $Z_\odot \leq Z \leq 0.3$ $Z_\odot$, which overlaps the range found in Galactic disk, thick disk and halo dwarf stars (Lauroesch *et al.* 1995).

Our understanding of the chemical evolution in the Galaxy is both guided and constrained by abundance measurements of halo, thick disk, disk, bulge and globular cluster stars. If most galaxies form in a similar fashion, the distinctive abundance patterns exhibited by Galactic stars should be discernible in the absorption line spectra produced by the gas which (accidentally) intercepts the line of sight of the background QSO. Conversely, the ensemble of QSO absorption systems may suggest a detailed model for the formation of the Galaxy which can be tested against the stellar abundance patterns. In §2, we discuss the coupling of various redshift–lookback time models to the abundance patterns observed in the Galaxy. These abundance patterns are then transformed into redshift space and compared to the abundances measured in QSO absorption line systems. Based on the agreement between the measured and derived redshift–abundance relations, it is suggested for elements not yet observed in QSO absorption line systems that the derived relations function as predictions. Extensions to the simple, first–order paradigm are examined and the resulting uncertainties thus introduced are discussed in §3. Conclusions are summarized in §4.

## 2. ELEMENT ABUNDANCES

The general expression for the lookback time – redshift relation is

$$H_o t = \int \frac{dz}{(1+z)\ E(z)} \quad , \tag{1}$$

where $H_o$ is the Hubble constant. The dimensionless function $E(z)$ is given by (see for example Peebles (1993))

$$E(z) = [\Omega(1+z)^3 + \Omega_R(1+z)^2 + \Omega_\Lambda]^{1/2} \quad , \tag{2}$$

subject to the constraint that
$$\Omega + \Omega_R + \Omega_\Lambda = 1 \ . \tag{3}$$

In this paper, attention is focused on cases where the cosmological constant is zero, so that the only free parameter is the density of matter $\Omega$. The above expressions are easy enough to integrate numerically; the results are shown in Figure 1 for two choices of the density parameter ($\Omega = 1.0$ and 0.2) and three choices of the Hubble constant ($H_o = 50$, 75 and 100 km sec$^{-1}$ Mpc$^{-1}$). For these parameters, the age of the universe (z=$\infty$) ranges from approximately 6 to 16 Gyr.

A first–order model, and perhaps the simplest model, for the QSO absorption line systems assumes that the absorbers follow an abundance history and dynamical evolution which is diffeomorphic [1] to the Galaxy (i.e., the Galaxy as a standard abundance candle; also see Wolfe 1987). This model is readily testable (and falsifiable!) by comparing the abundances observed in QSO absorption line systems with the abundance behavior, suitably transformed into redshift space, found in the Galaxy.

Knowledge of a Galactic age–metallicity relationship can be mapped into a redshift–metallicity relation, for a chosen cosmology, by using Figure 1. Hence, any abundance–metallicity trend can be transformed into an unambiguous abundance–redshift trend. To this transformation procedure a single free parameter is added, called the delay time. The delay time $\tau_{\rm delay}$ is the total time between the start of the Big Bang and onset of star formation. A more precise meaning of $\tau_{\rm delay}$ is given in §2.1 and in §3. This completely explains the simple first order model; one whose predictions, comparison to the observations and implications consume the remainder of this paper. It is important to recognize that we will be transforming observed abundance histories, not chemical evolution model results (e.g., Lanzetta, Wolfe & Turnshek 1995), into redshift space.

For completeness, the notation commonly used to describe the abundance of "X" relative to the solar value is
$$[X] \ = \ \log \frac{X}{X_\odot} \ = \ \log X - \log X_\odot \quad {\rm dex} \ . \tag{4}$$

In a few cases it is desirable to compare the abundance of "X" relative to hydrogen by number,
$$N(X) \ = \ \log \left(\frac{X}{H}\right) + 12.0 \quad {\rm dex} \ . \tag{5}$$

---

[1] Diffeomorphic, Diffeomorphism : A one-to-one differentiable mapping with a differentiable inverse. Primarily encountered in classical mechanics, relativistic mechanics, nonlinear dynamics and differential geometry.

## 2.1 AGE-METALLICITY-REDSHIFT RELATION

The small plot in the upper right hand corner of Figure 2 shows the age–metallicity (time versus [Fe/H]) relationship of the Galaxy inferred from spectral observations of dwarf stars in the solar vicinity (Edvardsson *et al.* 1993; Lambert 1989; Wheeler, Sneden & Truran 1989; Timmes, Woosley & Weaver 1995). This small plot constitutes the input. Note that it is scaled between 0 and 1 by dividing by the age of the universe, whatever value that may be for a chosen cosmology. Transposing the time coordinate into a redshift coordinate (see Figure 1) yields the main plot shown in Figure 2. Curves are shown for $\Omega=1.0$ and $\Omega=0.2$ cosmologies and three values of the delay, time $\tau_{\rm delay} = 0$, 1 and 3 Gyr. In this simple picture, no abundance evolution occurs between the start of the Big Bang and $\tau_{\rm delay}$. At $\tau_{\rm delay} + \delta$, the evolution shown in the upper right hand corner commences. The time required to reach a metallicity of [Fe/H] $\sim$ -1.0 dex, the canonical halo-disk transition period, is $\tau_{\rm delay}$ plus a few times $10^8$ years. A value of $\simeq 3 \times 10^8$ years is consistent with halo dynamical collapse timescales. In §3 we consider an alternative meaning for $\tau_{\rm delay}$.

Since it is required that the entire age–metallicity relationship conform to the age of the universe (minus any delay time), the redshift–metallicity evolutions are independent of the Hubble constant and depend only on $\Omega$. Although we will not explicitly state this fact for each group of elements that are considered below, it remains true for all of the redshift–abundance evolutions.

Spectroscopic abundance determinations of the zinc to hydrogen ratio [Zn/H] in damped Lyman-$\alpha$ absorption line systems are shown in Figure 2 as the solid circles, and represent a compilation of measurements obtained by Meyer, Welty & York (1989), Meyer & Roth (1990), Meyer & York (1992), Savaglio, D'Odorico & Møller (1994), Pettini *et al.* (1994), Wolfe *et al.* (1994), Kulkarni *et al.* (1995), York, Welty & Meyer (1995), and Lauroesch (1995). The typical error in determining the redshift is much smaller than the solid circles, while the typical errors (not including systematic uncertainties like ionization corrections) have been estimated to be be $\sim 20\%$ for hydrogen and 10–20% for zinc.

If one is to compare the [Zn/H] observations with the [Fe/H] calculations, two conditions must be met. First, zinc and iron must track each other, i.e [Zn/Fe] = 0, over the entire metallicity range. Several high resolution, low noise surveys that employ digital spectra of the Zn I lines at 4722 Å and 4810 Å have shown that the zinc to iron ratio in Galactic halo and disk stars is indeed solar, with a very small star-to-star scatter over the metallicity range -2.9 dex $\leq$ [Fe/H] $\leq$ -0.1 dex (Sneden & Crocker 1988; Sneden, Gratton & Crocker 1991; Ryan, Norris & Bessell 1991). For example, Sneden *et al* (1991) in their study of 23 field dwarfs found [Zn/Fe] = 0.04 $\pm$ 0.01 dex, and that this ratio result is robust with respect to variations in the effective temperatures and surface gravities. Stars in the LMC and SMC also display [Zn/Fe] $\simeq$ 0 (Russell & Dopita 1992). Theoretical calculations of stellar nucleosynthesis and chemical evolution also suggest that zinc and iron are produced in tandem over the entire metallicity range (Woosley & Weaver 1995; Thielemann, Nomoto & Hashimoto 1995; Arnett 1995; Timmes *et al.* 1995).

The second condition to be met is that zinc should not be significantly depleted onto dust grains. Extensive surveys of the interstellar medium have shown that zinc, relative to hydrogen, is generally present in the gas phase in nearly solar proportions, and that it is depleted onto dust by at most a factor of 2–3 in the densest clouds surveyed (Morton 1975; Shull, York & Hobbs 1977; De Boer *et al.* 1986; Jenkins 1987; Van Steenberg & Shull 1988; Cardelli *et al.* 1991; Sembach *et al.* 1995). Thus, zinc is not readily incorporated onto dust grains. Hence, it appears justifiable to compare the [Zn/H] observations with the [Fe/H] calculations of Figure 2. Iron in the gas phase tends to be depleted onto grains, so any direct comparison of an iron abundance measured QSO absorption line systems will have (large) depletion correction factors. The main conclusions to be drawn from Figure 2 are examined below, after Figures 3 and 4 are discussed.

We have considered a single-valued function for the input age–metallicity relationship (small plot in Fig. 2). The observed age-metallicity relation for disk stars (Edvardsson *et al.* 1993), halo stars (Schuster & Nissen 1989) and globular clusters (VandenBerg, Bolte & Stetson 1990; Zinn 1993) all strongly suggest that there is an intrinsic spread in metallicity for a given age (or vice versa, a spread in age for a given metallicity). The net result of including an intrinsic spread would be to broaden the crisp lines of Figure 2 into bands. It is possible, although not examined or proven in any detail in this paper, that the width of these bands could adequately explain almost all of the scatter in the QSO absorption line abundances.

The lone data point in Figure 2 with a redshift < 1.0 is from 3C 286 (Meyer & York 1992). Suprisingly they found that the zinc abundance ([Zn/H] = -1.2 dex) was the same as z ∼ 2.0 QSO absorption line systems. Steidel *et al.* (1994) suggested that the absorber in front of 3C 286 is a low surface brightness galaxy ($\mu_B(0) \simeq 23.6$ mag arcsec$^{-2}$). Chemical enrichment in low surface brightness galaxies apparently proceeds at a slower pace in these systems, even in the present epoch, than in "normal" spirals (Pettini *et al.* 1995). This anamolous data point in Figure 2 indicates that there is probably some fraction of QSO absorbers that do not fit into our simple paradigm.

## 2.2 $\alpha$–CHAIN ABUNDANCE HISTORIES

Evolution of the $\alpha$–chain abundances (oxygen, neon, magnesium, silicon, sulfur, argon, calcium and titanium) are shown in Figure 3. The small plot in the upper right hand corner forms the basic input and represents a schematic evolution of the $\alpha$–chain elements with the metallicity [Fe/H] as inferred from surveys of solar neighborhood stars (Spite &

Spite 1985; Wheeler *et al.* 1989). The precise form of this evolution is, of course, slightly different for each of the different elements. However, $\alpha$–chain elements generally follow the pattern indicated by the small plot: an enhancement by roughly a factor of three for [Fe/H] $\lesssim$ -1.0 dex, followed by a linear decline to solar values in the range -1.0 dex $\lesssim$ [Fe/H] $\lesssim$ 0.0 dex, with perhaps a smaller change in slope at enhanced metallicities [Fe/H] $\gtrsim$ 0.0 dex (Lambert 1989; Wheeler *et al.* 1989). Shifting the metallicity dependence of the small plot into redshift space yields the main plot of Figure 3, where the $\alpha$–chain abundances, relative to hydrogen, are shown as a function of redshift for $\Omega$=1.0 and $\Omega$=0.2 cosmologies and three values of the delay time $\tau_{\text{delay}}$ = 0, 1 and 3 Gyr.

Spectroscopic abundance determinations of the silicon to hydrogen ratio [Si/H] in the damped Lyman–$\alpha$ systems are shown in Figure 3 as the solid circles. The observations are a compilation of observations by Blades it et al. 1985, Meyer, Welty & York (1989), Pettini & Hunstead (1990), Rauch *et al.* (1990), Carswell *et al.* (1991), Savaglio, D'Odorico & Møller (1994), Kulkarni *et al.* (1995), York, Welty & Meyer (1995), Pettini, Lipman & Hunstead (1995), and Lauroesch (1995). All of the silicon abundance determinations utilized the observed Si II column densities, and the ionization corrections made by Carswell *et al.* (1991) and Lauroesch (1995) were removed to put all the measurements on an equal footing. Typical redshift errors in these [Si/H] determinations are much smaller than the solid circles, while the uncertainties in the abundances have been estimated to be $\sim$ 20% for hydrogen and 10–20% for silicon. Silicon probably has similar or perhaps larger grain depletion factors than zinc, but these depletion corrections only shift the observations in the vertical direction.

It can be helpful in understanding the abundance trends and redshift transformations to examine the $\alpha$–chain abundances relative to iron instead of hydrogen. Figure 4 shows this change of basis and clearly illustrates the trend of increasing [$\alpha$/Fe] ratios as a function of redshift. The small "glitches" at the top of the plot are manifestations of the sharp change in [$\alpha$/Fe] slope that was input at [Fe/H] = -1.0 dex (see the small inset plot in the lower right hand corner of Figure 4) and are not important.

## 2.3 INTERLUDE

The main conclusion to be drawn from Figures 2 and 3 is that all of the [Zn/H] and [Si/H] damped Lyman–$\alpha$ observations are consistent with a delay time between the start of the Big Bang and the formation of a thin disk of scale of several Gyr, for all reasonable choices of $\Omega$ in a $\Lambda$=0 cosmology. We stress that in this simple first–order model, no abundance evolution occurs between the start of the Big Bang and $\tau_{\text{delay}}$. An alternative, more complicated interpretation of $\tau_{\text{delay}}$ in given in §2.6. That the simple, first–order model for the chemical histories of the QSO absorption line systems fits the zinc and silicon observations over more than two orders of magnitude in abundance is surprising and encouraging.

However, prudent caution is still advisable. The large scatter in the abundances and the narrow window of redshifts over which they have been observed (1.8 < z < 2.3) could conspire to give the illusion that the chemical enrichment began at these same redshifts (Fall 1995, private communication). There may be selection effects such as dust obscuration (Pei & Fall 1995) or significant line of sight variations that could substantially broaden the present redshift window. If this turns out to be the case, then our conclusion about a few Gyr hiatus before abundance evolution commences may have to be abandoned, or at least modified.

## 2.4 ODD-Z NUCLEI EVOLUTIONS

Abundance histories of several elements with an odd number of protons in the nucleus (sodium, aluminum, chlorine, phosphorus, manganese, copper) are shown in Figure 5. The small plot in the upper right corner constitutes the input and schematically shows the evolution of these elements with metallicity, as suggested by abundance determinations of Galactic halo and field dwarfs. Details of each element's evolution with [Fe/H] can be different from the schematic shown in Figure 5 (in particular aluminum and copper), but each element *generally* follows the pattern of being subsolar by roughly a factor of 3 in the metallicity range -3.0 dex $\lesssim$ [Fe/H] $\lesssim$ -1.0 dex, followed by a roughly linear increase to solar values at [Fe/H] $\simeq$ 0.0 dex (Lambert 1989; Wheeler *et al.* 1989). The fluorine to oxygen ratio [F/O] may also be included in this group of elements or, due to uncertainties in the neutrino process yields, may more appropriately be included in the group of elements examined in Figure 6.

Transforming the input abundance trends into redshift space for $\Omega$=1.0 and $\Omega$=0.2 cosmologies and three values of the delay time $\tau_{\text{delay}}$ = 0, 1 and 3 Gyr yields the main plot of Figure 5. Systematic studies of the abundances of these elements have not yet been performed for QSO absorption line systems, so that abundance patterns of Figure 5 serve as first–order predictions. Several elements in this group may have large grain depletion factors (e.g. aluminum, calcium and manganese) that would affect placement of the observed QSO absorption line system abundances in the figure.

It is somewhat difficult to decide if potassium should be grouped in with the odd-Z nuclei or with the $\alpha$–chain elements discussed above. Gratton & Sneden's (1987) survey of potassium abundances in 23 northern and southern hemisphere stars, which included both dwarfs and field giants, suggested that [K/Fe] $\sim$ 0.0 dex in disk stars, while [K/Fe] $\geq$ 0.0 dex in metal-poor stars but with a very large scatter. They noted that the K I resonance lines are often heavily blended with extremely strong lines of atmospheric molecular oxygen and that the low excitation energies of the K I lines may be strongly susceptible to overionization or strong hyperfine structure effects. Thus, it is possible that [K/Fe] may actually be $\lesssim$ 0.0 dex in halo dwarfs (see Timmes *et al.* 1995).

## 2.5 IRON CO-PRODUCTION ELEMENTS

Evolution of carbon, nitrogen, scandium, vanadium, chromium, cobalt, nickel and zinc are shown in Figure 6. The small plot in the upper right hand corner is based on spectroscopic surveys of stars in the Galactic halo and disk, where one expects the principle trend of these elements, relative to iron, to be flat and solar for all metallicities. Superimposed on this general trend (shown as the shaded regions in Figure 6) there is some intrinsic scatter, which for some elements (e.g carbon) may be as large as $\pm$ 0.2 dex. The precise degree of flatness and intrinsic scatter is, of course, unique to each element. The fluorine to oxygen ratio [F/O] may considered be a member of this group of elements, rather than the elements discussed in Figure 5. Transforming the input yields no surprises in the main plot, which shows the element abundances relative to iron as a function of redshift. Thus, one expects the evolution of these elements, relative to hydrogen, to be identical to those shown in Figure 2, modulo any intrinsic scatter.

## 2.6 LITHIUM, BERYLLIUM AND BORON

Abundances histories of lithium and boron, relative to hydrogen by number, are shown in Figure 7. The plot inset in the upper right hand corner of Figure 7 schematically shows the evolution of these two elements with metallicity, as dictated by abundance determinations of Galactic halo and field dwarfs.

Stellar lithium abundance determinations that use high resolution, low noise digital spectra of the Li I 6707 Å doublet (Spite & Spite 1982; Boesgaard & Tripico 1986; Rebolo, Molaro & Beckman 1988; Thorburn 1994) suggest that, between -3.0 dex $\leq$ [Fe/H] $\leq$ -1.0 dex, N(Li)$\simeq$2.2 with a very small star-to-star scatter (i.e the Spite plateau). At [Fe/H] $\simeq$ -1.0 dex, the upper envelope of the observations smoothly increases to a maximum value that is an order of magnitude greater than those found in the most metal-poor dwarfs. It is common practice to assume that the lithium abundance of very metal-poor stars represents the primordial value. An alternative suggestion, based on calculations of inhomogeneous Big Bang nucleosynthesis, is that the primordial value is actually much higher, N(Li) $\gtrsim$ 3, and all the values measured today are the results of depletion (Alcock, Fuller & Mathews 1987; Malaney & Fowler 1988; Fuller, Mathews & Alcock 1988; Applegate, Hogan & Scherrer 1988; Brown 1992). For the purposes of mapping the lithium abundances into redshift space, we will assume that the Spite plateau represents the primordial $^7$Li abundance.

Boron abundance determinations that employ high resolution, low noise digital spectra of the B I resonance line near 2497 Å, corrected for various departures from local thermodynamic equilibrium (Duncan, Lambert & Lemke 1992; Lemke, Lambert & Edvardsson 1993; Kiselman 1994; Edvardsson *et al.* 1994), strongly suggest a linear relationship between boron abundances and the metallicity [Fe/H]. Beryllium abundances determined in halo stars through high resolution, digital spectra of the Be II 3131 Å resonance doublet (Rebolo *et al.* 1989; Ryan *et al.* 1990; Gilmore, Edvardsson & Nissen 1991; Gilmore *et al.* 1992; Ryan *et al.* 1992) indicate that beryllium is also linear with [Fe/H]. The Be abundances tend to be smaller than boron abundances by roughly a factor of ten, and the slope with of the beryllium abundances with [Fe/H] is slightly larger than the boron-[Fe/H] slope.

Transformation of the lithium and boron abundances into redshift space yields the main plot of Figure 7. The expected abundance patterns are shown for $\Omega=1.0$ and $\Omega=0.2$ cosmologies and three values of the delay time $\tau_{\text{delay}} = 0$, 1 and 3 Gyr. The berryllium abundances discussed above have not been transformed into redshift space because, to first order, the beryllium abundances are the boron abundances decreased by $\sim 1.0$ dex. Since none of these light elements has yet been observed in QSO absorption line systems, the calculated abundances of Figure 7 serve as guideline predictions.

## 3.0 PANDORA'S BOX

The redshift–abundance patterns given in the previous sections used a delay time which was interpreted to mean that no abundance evolution occurred between the start of the Big Bang and $\tau_{\text{delay}}$. This is the simplest and most unambiguous case. Any other interpretation or modification, while yielding the next order correction to the model, opens a Pandora's box of complications and uncertain parameters. As an example, we consider the case where the total time delay $\tau_{\text{delay}}$ is the sum of a no abundance evolution phase and a halo phase. The halo phase is characterized by how long ($\tau_{\text{halo}}$) a classical halo dwarf abundance pattern persists. For times larger than $\tau_{\text{delay}}$, the abundance evolution is taken to follow the paths they did in the previous sections.

Values $\tau_{\text{halo}}$ which are several Gyr appear to agree quite well with several independent lines of reasoning:

(1) The age of the Galactic halo population of globular clusters is estimated to lie between 13–17 Gyr (see the review by Cowan, Thielemann & Truran 1991). This age is generally assumed to reflect the age of the Galaxy itself, as the spheroidal population of globular cluster certainly dates the time of the halo collapse. In contrast, the disk component of the Galactic globular cluster system is characterized by very different kinematics and metallicity distribution, and appears to be a few Gyr younger (Zinn 1985; Lee, Demarque & Zinn 1990).

(2) VandenBerg, Bolte & Stetson (1990) suggested that while the most metal deficient globular clusters ([Fe/H] $\simeq$ -2.1 dex) appear to be coeval, the metal rich halo systems ([Fe/H] $\simeq$ -1.3 dex) show a spread in age of a few Gyr. Based upon their distinctive horizontal branch morphologies, Zinn (1993) also suggested that the halo globular cluster population can be divided into two groups that appear to constitute two distinct age populations.

(3) Schuster & Nissen (1989) identified an age–metallicity relationship for the halo stars, with an inferred age spread of a few Gyr. The age–metallicity and velocity dispersion–age relationships of the Edvardsson *et al.* (1993) survey indicate distinct transitions at a metallicity between -1.0 dex and -1.1 dex, with the implied time scale and stellar ages consistent with a several Gyr hiatus before formation of the thin disk.

(4) The young age determined for the disk population of white dwarfs (Winget *et al.* 1987; Iben & Laughlin 1989) seems consistent with a several Gyr halo timescale, although here one must be careful to include a nonzero timescale for the cooling of the disk and the associated restriction of star formation to a thin disk region (Burkert, Truran & Hensler 1992).

(5) One dimensional hydrodynamical models of the Galaxy's halo-to-disk evolution suggest that the energy deposited by Type II supernovae delays the appearance of the thin disk by several Gyr (Burkert, Truran, & Hensler 1992).

(6) Chemical evolution models that use the infall of primordial or near primordial gas on few Gyr *e*-folding time scale tend to achieve a reasonable simultaneous fit to the solar vicinity gas surface density and total surface mass density (stars + gas), and to the abundance histories of the elements (Chiosi 1980; Chiosi & Matteucci 1982; Wheeler *et al.* 1989)

The redshift-metallicity relation for $\tau_{\text{delay}} = 3.0$ Gyr (see Figure 2), is shown in Figure 8 for $\tau_{\text{halo}} = 0$, 1.0, 2.0 and 3.0 Gyr. Only the $\tau_{\text{halo}} = 1.0$ Gyr and 3.0 Gyr curves, for $\Omega$=1.0, are labeled in the figure. The difference between $\tau_{\text{delay}}$ and $\tau_{\text{halo}}$ is the time period where no abundance evolution takes place. Bifurcation of the curves at [Fe/H] = -1.0 occurs because that metallicity was artificially selected to mark the transition point between the halo and disk abundance evolutions. Unlike the simple, first–order model where the abundance patterns were independent of the Hubble constant, setting $\tau_{\text{halo}}$ breaks the scale-free property and causes the Hubble constant choice to affect the evolutions. All the curves shown in Figure 8 are for a Hubble constant of $H_o = 75$ km sec$^{-1}$ Mpc$^{-1}$, with other values simply translating the curves horizontally.

Unlike the abundance evolutions of Figure 2, the abundance evolutions shown in Figure 8 have the metallicity pleasantly decreasing with redshift. The expense of this improvement in the simple first–order model is the opening of a Pandora's box: the time alloted to the no evolution phase and the halo phase must be partitioned, a disk-halo transition point must be selected, and the evolutions now depend on the value of the Hubble constant. To turn the situation around, as the number of high quality abundance determinations in QSO absorption line systems increase, perhaps they will be able to distinguish between the uncertainties and assist in selecting the correct parameters for extensions to the simple first–order paradigm.

## 4. CONCLUSIONS

Abundance histories of the QSO absorption line systems as a function of redshift for metals lighter than gallium have been calculated for various cosmological models coupled to a simple, first–order model for the chemical evolution of these systems. The first order model assumes that the QSO absorption line systems follow an abundance history and dynamical evolution that is diffeomorphic to the Galaxy (i.e the Galaxy as a standard abundance candle), which allow transformation of the observed abundance trends in the Galaxy to abundance histories for the gas in QSO absorption line systems. Comparison of the transformed abundance patterns with the observed zinc to hydrogen [Zn/H] and silicon to hydrogen [Si/H] ratios in QSO absorption line systems found the observations agreed with a $\Lambda=0$, $\Omega=0.2 - 1.0$, $\tau_{\rm delay}=3$ Gyr model over more than two orders of magnitude in abundance. For elements which have not yet been observed in QSO absorption line systems, the transformed abundances, modulo intrinsic scatter and grain depletion factors, serve as first-order predictions. Extensions to the simple first–order paradigm were shown to give a tailing off of the abundances with redshift, at the expense of introducing uncertain parameters.

This work has been supported by a grant from the National Science Foundation (AST92-17969), a grant from NASA (NAG 5-2770), and an Enrico Fermi Postdoctoral Fellowship (FXT).

**Figures and Captions**

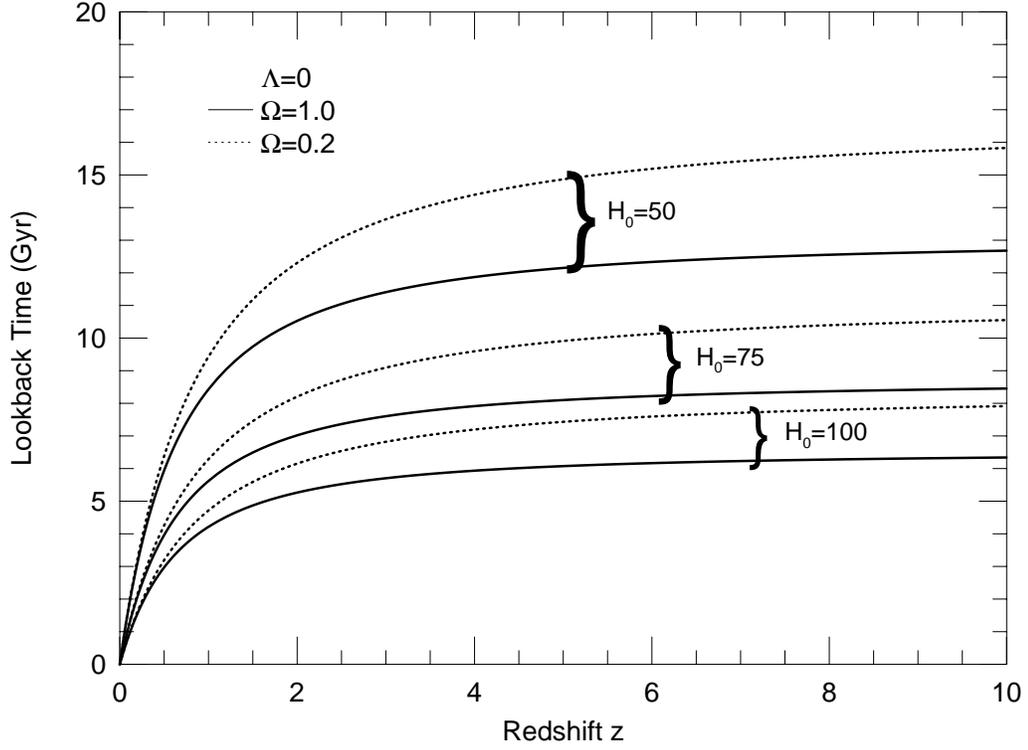

Fig. 1.— Lookback time as a function of redshift for the case when the cosmological constant can be ignored. The only free parameter is then the density of matter parameter $\Omega$. The figure shows $\Omega = 1.0$ and $\Omega = 0.2$ cosmologies for three choices of the Hubble constant; $H_o = 50$, 75 and 100 km sec$^{-1}$ Mpc$^{-1}$. For these parameters, the age of the universe (z=$\infty$) ranges from approximately 6 to 16 Gyr.

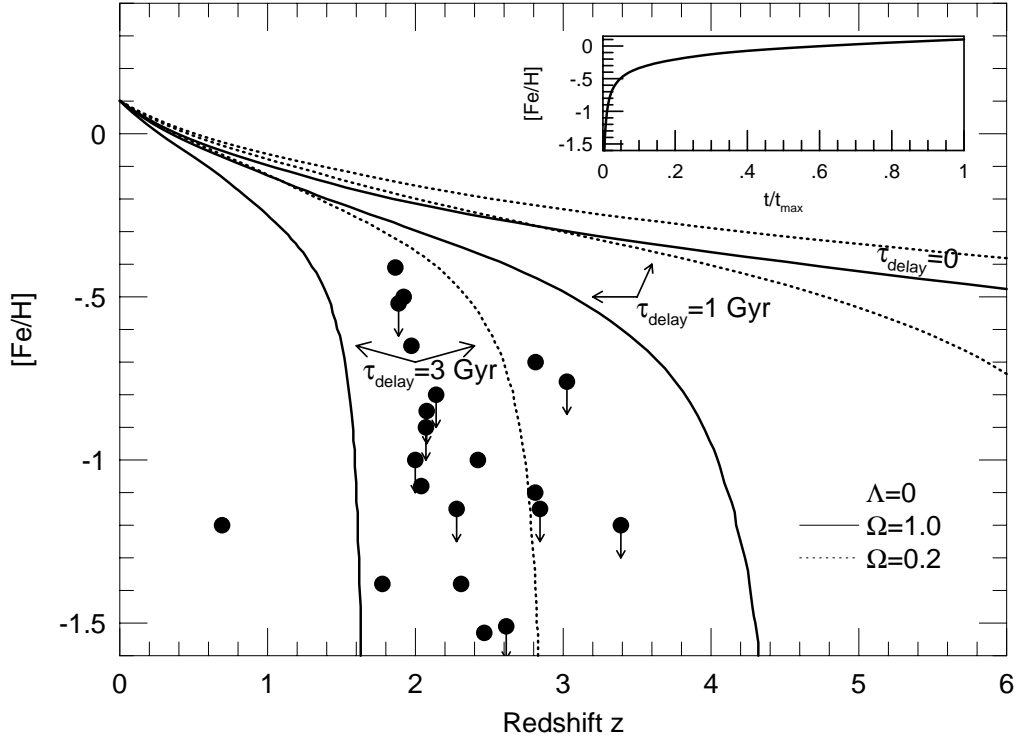

Fig. 2.— Iron to hydrogen ratio evolutions. The small plot in the upper right hand corner is the age–metallicity relationship of the Galaxy as inferred from spectral observations of nearby dwarf stars and constitutes the input. Note that the x-axis in the small plot is scaled so it is independent of the age of the universe. The main plot shows the output of the first-order model for $\Omega=1.0$ and $\Omega=0.2$ cosmologies and three values of the delay time $\tau_{\rm delay} = 0$, 1 and 3 Gyr. Spectroscopic abundance determinations of the zinc to hydrogen ratio [Zn/H] in the QSO absorption line systems are shown in the figure as solid circles (see text for discussion). For all reasonable choices of $\Omega$ in a $\Lambda=0$ cosmology, the observations are consistent, for more than two orders of magnitude in abundance, with 3 Gyr delay time.

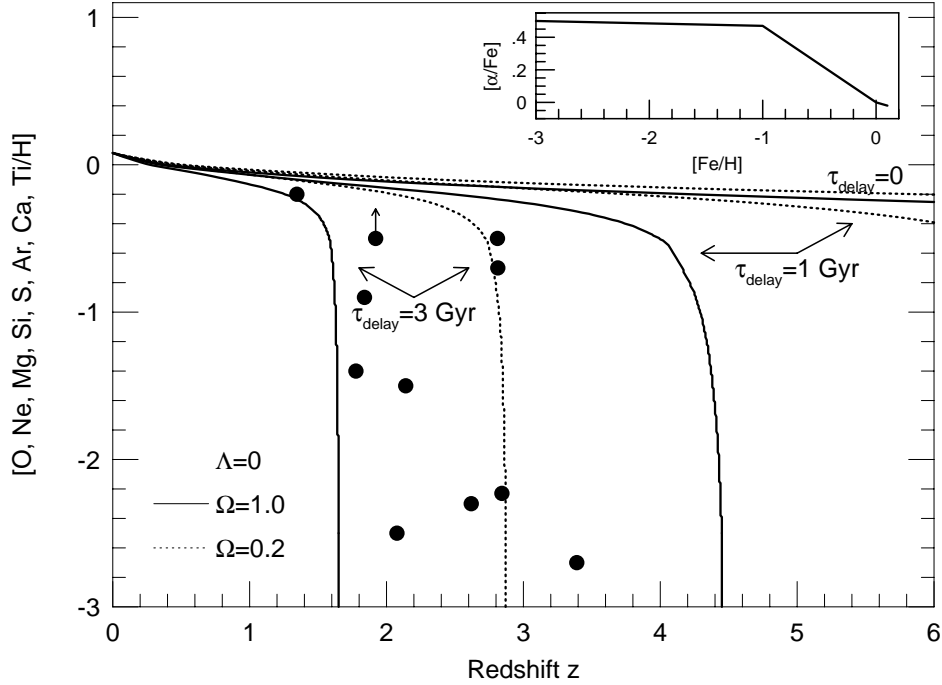

Fig. 3.— Abundance evolutions of oxygen, neon, magnesium, silicon, sulfur, argon, calcium and titanium. The small plot in the upper right hand corner is the input and schematically represents the observed evolution of the $\alpha$–chain elements with [Fe/H], as inferred from surveys of solar neighborhood stars. The main plot is the output and shows the $\alpha$–chain abundances, relative to hydrogen, as a function of redshift for $\Omega=1.0$ and $\Omega=0.2$ cosmologies and three values of the delay time $\tau_{\text{delay}} = 0$, 1 and 3 Gyr. Spectroscopic abundance determinations of the silicon to hydrogen ratio [Si/H] in QSO absorption line systems are shown as the solid circles (see text for reference list). Comparison of the observed silicon abundances to the simple, first order model abundances suggests that, for all reasonable choices of $\Omega$ in a $\Lambda=0$ cosmology, the observations are consistent, over more than 2 orders of magnitude in abundance, with 3 Gyr delay time.

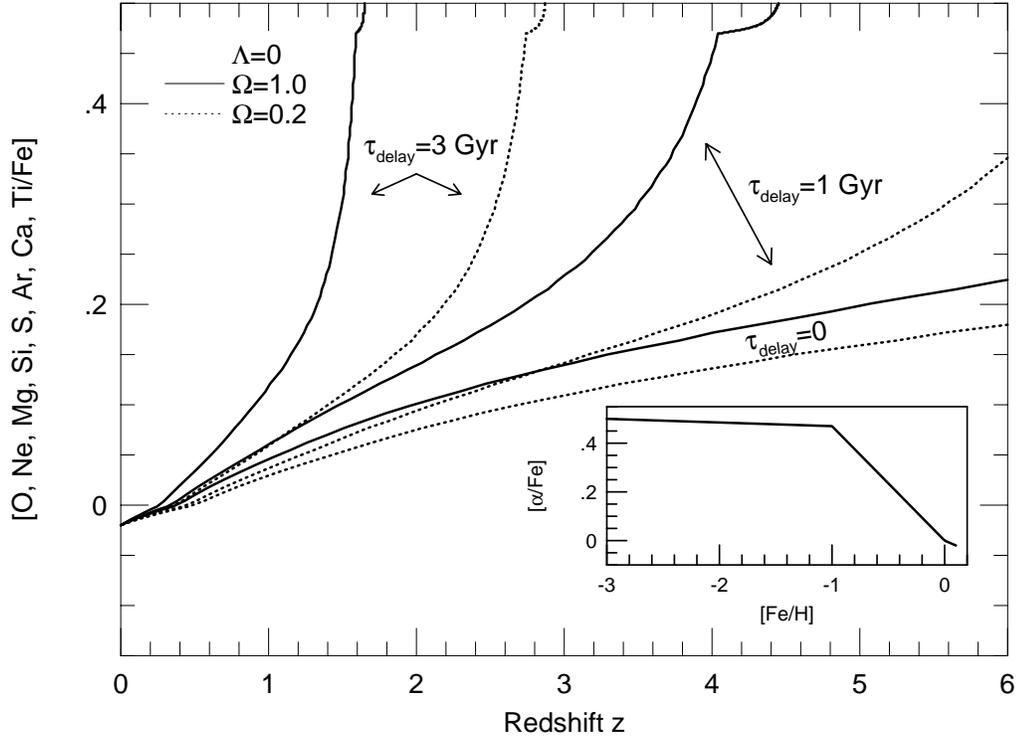

Fig. 4.— $\alpha$–chain evolutions relative to iron as a function of redshift. The small "glitches" at the top of the plot are manifestations of the sharp change in slope that was input at [Fe/H] = -1.0 dex

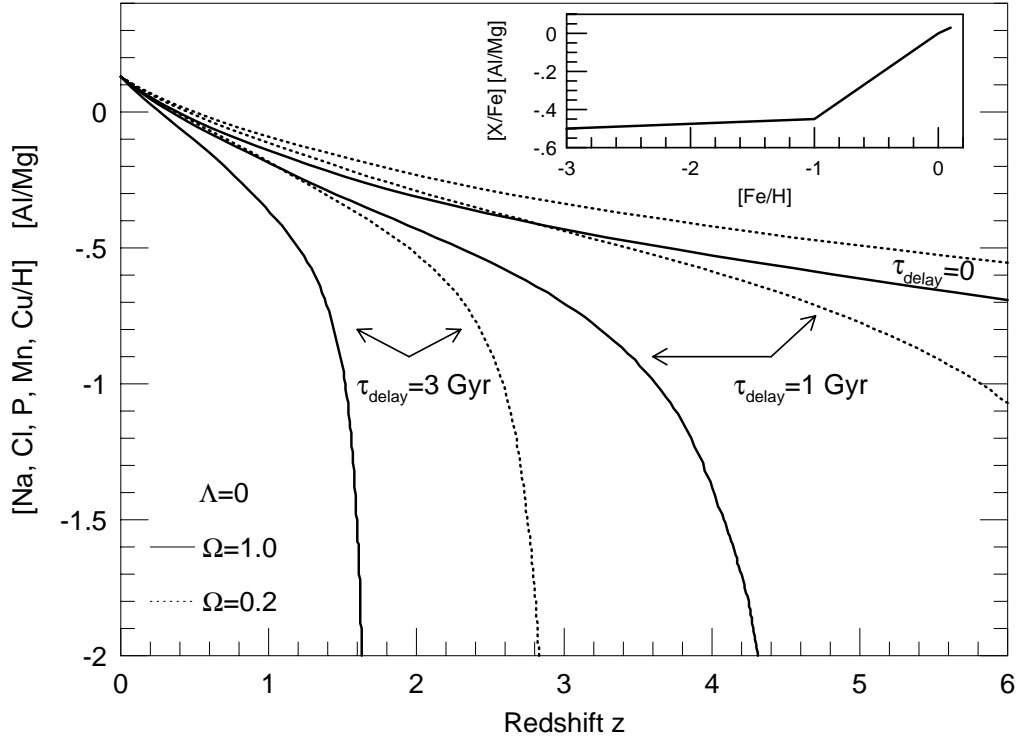

Fig. 5.— Abundance histories of sodium, aluminum, chlorine, phosphorus, manganese, copper. The fluorine to oxygen ratio [F/O] may be included in the group of elements. The small plot in the upper right corner constitutes the input and schematically shows the evolution of these elements with metallicity, as suggested by abundance determinations of Galactic halo and field dwarfs. Transforming the input abundance trends into redshift space yields the expected abundance patterns shown in the main plot for $\Omega$=1.0 and $\Omega$=0.2 cosmologies and three values of the delay time $\tau_{\rm delay} = 0$, 1 and 3 Gyr.

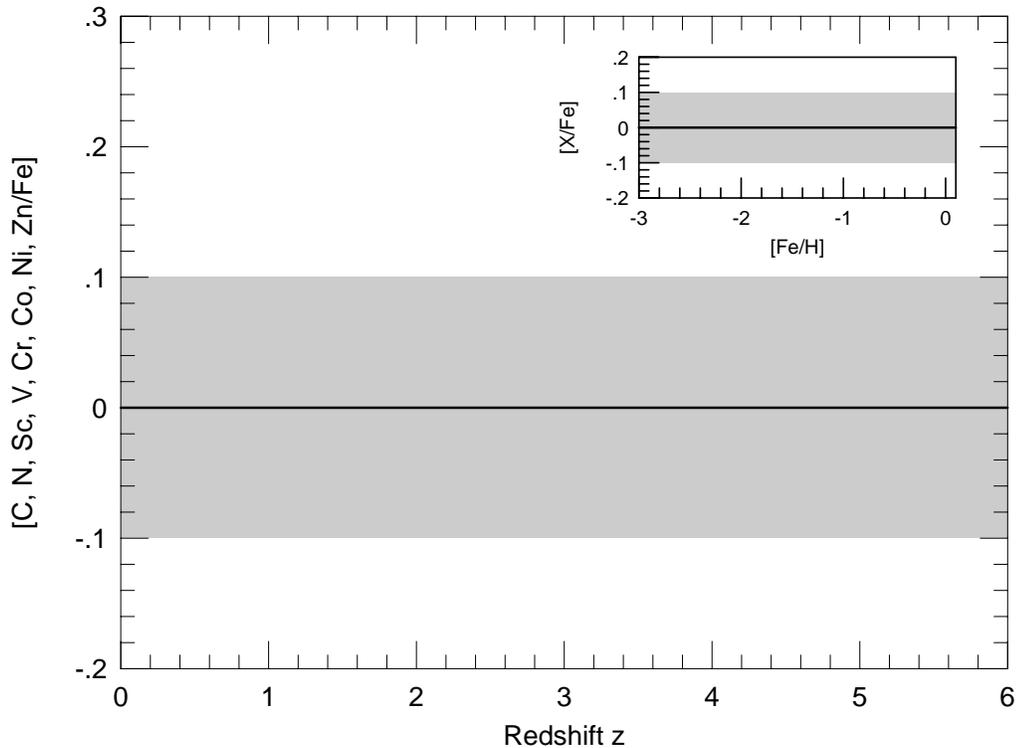

Fig. 6.— Evolution of carbon, nitrogen, scandium, vanadium, chromium, cobalt, nickel and zinc, relative to iron. The small plot in the upper right hand corner is based on spectroscopic surveys of stars in the Galactic halo and disk, where one expects the principle trend of these elements, relative to iron, to be flat (solar) for all metallicities. The shaded region schematically shows the size of the intrinsic scatter encountered in stellar abundance determinations. The precise degree of flatness and intrinsic scatter is, of course, unique to each element. The fluorine to oxygen ratio [F/O] may belong to this group of elements. Transforming the input yields the main plot, which shows no surprises in the abundance trends with redshift. Evolution of these elements relative to hydrogen should identical to Figure 2, modulo any intrinsic scatter and grain depletion effects.

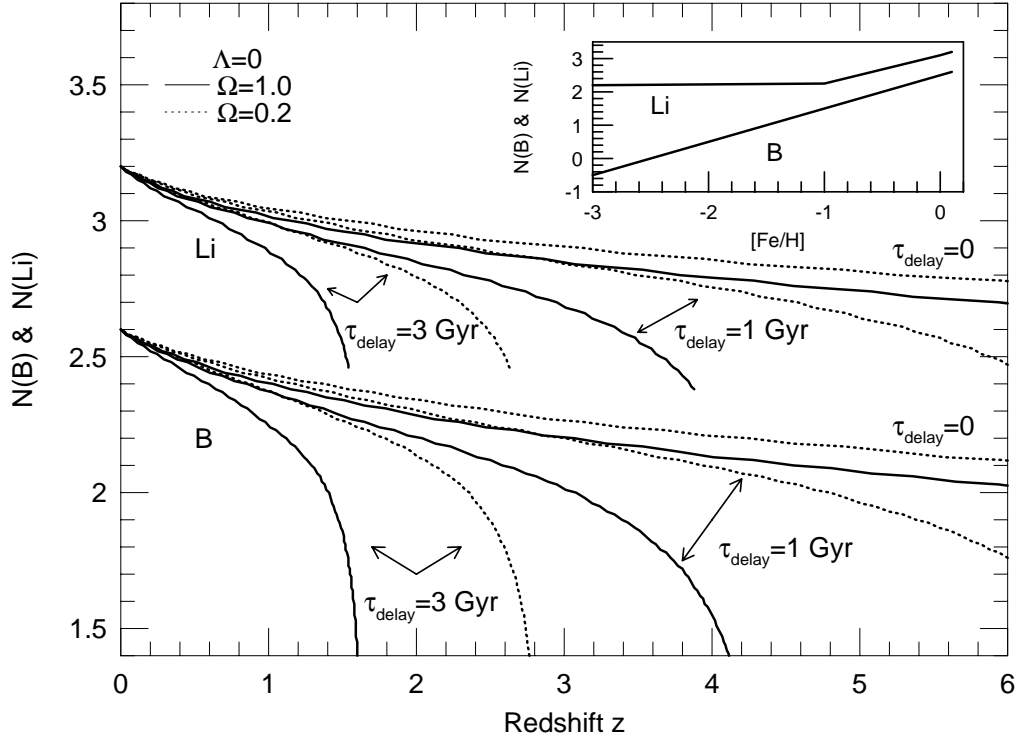

Fig. 7.— Abundance histories of lithium and boron. The small plot in the upper right hand corner schematically indicates the evolution of these two elements, relative to hydrogen by number, with metallicity, as suggested by the abundance determinations of Galactic halo and field dwarf stars. For the purposes of mapping the lithium abundances into redshift space, it is assumed that the lithium abundances of the most metal poor halo dwarfs represent the primordial $^7$Li abundance. The main plot shows the expected abundance patterns for $\Omega$=1.0 and $\Omega$=0.2 cosmologies and three values of the delay time $\tau_{\text{delay}} = 0$, 1 and 3 Gyr.

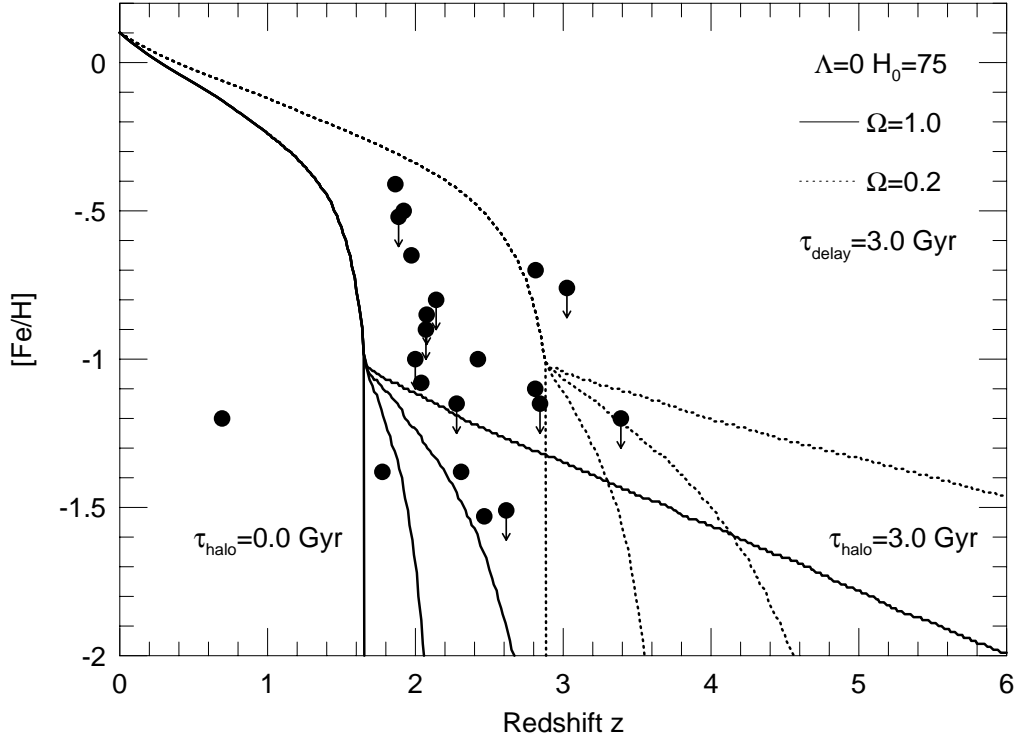

Fig. 8.— Redshift-metallicity relations for a time delay of 3.0 Gyr and halo timescales of 0.0, 1.0, 2.0 and 3.0 Gyr. Only the $\Omega$=1.0, $\tau_{halo}$ = 1.0 Gyr and 3.0 Gyr curves are labeled. The difference between $\tau_{delay}$ and $\tau_{halo}$ is the time period where no abundance evolution takes place. The transition point between the halo and disk abundance evolutions was artificially set at [Fe/H] = -1.0 dex, which causes all the curves to bifurcate at that metallicity. All the curves shown are for a Hubble constant of $H_o = 75$ km sec$^{-1}$ Mpc$^{-1}$, with other values translating the curves horizontally. Improving the simple first-order paradigm shown in Figure 1 carries the price of opening a Pandora's box of uncertain parameters (see text).